\documentclass[twocolumn,english,aps,prl,floats,showpacs]{revtex4}
\usepackage[T1]{fontenc}
\usepackage[latin9]{inputenc}
\usepackage{amsmath}
\usepackage{amssymb}
\usepackage{graphicx}
\usepackage{esint}

\makeatletter

\newcommand{\lyxdot}{.}

\@ifundefined{textcolor}{}
{%
 \definecolor{BLACK}{gray}{0}
 \definecolor{WHITE}{gray}{1}
 \definecolor{RED}{rgb}{1,0,0}
 \definecolor{GREEN}{rgb}{0,1,0}
 \definecolor{BLUE}{rgb}{0,0,1}
 \definecolor{CYAN}{cmyk}{1,0,0,0}
 \definecolor{MAGENTA}{cmyk}{0,1,0,0}
 \definecolor{YELLOW}{cmyk}{0,0,1,0}
}

\usepackage{babel}

\makeatother

\usepackage{babel}
\begin{document}

\title{Random Fields, Topology, and The Imry-Ma Argument}

\author{Thomas C. Proctor, Dmitry A. Garanin, and Eugene M. Chudnovsky}

\affiliation{Physics Department, Lehman College, City University of New York \\
 250 Bedford Park Boulevard West, Bronx, New York 10468-1589, USA}

\date{\today}
\begin{abstract}
We consider $n$-component fixed-length order parameter interacting
with a weak random field in $d=1,2,3$ dimensions. Relaxation from
the initially ordered state and spin-spin correlation functions have
been studied on lattices containing hundreds of millions sites. At $n-1<d$
presence of topological structures leads to metastability, with the
final state depending on the initial condition. At $n-1>d$, when
topological objects are absent, the final, lowest-energy, state is independent of
the initial condition. It is characterized by the exponential decay of 
correlations that agrees quantitatively with the theory based upon the
Imry-Ma argument. In the borderline case of $n-1=d$, when topological
structures are non-singular, the system possesses a weak metastability
with the Imry-Ma state likely to be the global energy minimum. 
\end{abstract}

\pacs{05.50.+q, 64.60.De, 75.10.Nr}

\maketitle
Almost forty years ago Imry and Ma argued \cite{Imry-Ma-PRL1975}
that for a model described by the Hamiltonian 
\begin{equation}
{\cal H}=\int d^{d}r\left[\frac{\alpha}{2}(\nabla{\bf S})^{2}-{\bf h}\cdot{\bf S}\right]\label{Ham-continuous}
\end{equation}
the random field, ${\bf h}({\bf r})$, regardless of its strength
$h$, destroys the long-range order associated with the continuous-symmetry
$n$-component order parameter ${\bf S}$ below $d=4$ spatial dimensions.
In the resulting Imry-Ma state there is only short-range ordering
within randomly oriented (Imry-Ma) domains of average size $R_{f}$
which depends on the strength of the random field. A similar argument had been made earlier
by Larkin \cite{Larkin-JETP1970} for pinned flux lattices in superconductors. 
Aizenman and Wehr published a rigorous analytical proof of the destruction
of the long-range order by the random field \cite{Aizenman-Wehr}.
Countless papers applied the Imry-Ma argument to random magnets \cite{EC-PRB86},
arrays of magnetic bubbles \cite{bubbles}, superconductors \cite{Blatter-RMP1994},
charge-density waves \cite{Efetov-77}, liquid crystals \cite{LC},
etc. Soon, however, the picture of Imry-Ma domains became subject
of a significant controversy. Theoretical work based upon renormalization
group and replica-symmetry breaking methods questioned the validity
of the argument for distances beyond $R_{f}$ \cite{Nattermann-2000}.
Numerical work \cite{Dieny-PRB1990,DC-PRB1991,Gingras-Huse-PRB1996,Fisch-PRB07,GCP-EPL13,GCP-PRB13} added to the controversy by revealing glassy properties of the random-field
model: The final state seemed to depend on the initial condition,
making the statement about the destruction of the long-range order
meaningless. It was suggested in Ref.\ \onlinecite{GCP-PRB13}
that the high energy cost of vortices in two dimensions and vortex
loops in three dimensions was preventing the spins in the $xy$ random-field
model from relaxing to a disordered state from the initially ordered
state.

In this Letter we show more generally that the long-range behavior
of random-field systems is controlled by topology. The latter, for
the model described by the Hamiltonian (\ref{Ham-continuous}) with $n$-component ${\bf S}$ and $n$-component ${\bf h}$, depends on the relation between $n$ and $d$. The condition ${\bf S}^{2}=S_0^2={\rm const}$
leaves $n-1$ components of the field independent. At $n-1<d$, mapping
of $n-1$ independent parameters describing the field ${\bf S}$ onto
$d'<d$ spatial dimensions provides topological defects with singularities.
They are vortices in the $xy$ model ($n=2$) in 2$d$, vortex loops
in the $xy$ model in 3$d$, and hedgehogs in the Heisenberg model
($n=3$) in 3$d$. Energy barriers associated with creation/annihilation
of these topological defects and their pinning by the random field
make the final state of the system strongly dependent on the initial
condition, thus invalidating the Imry-Ma argument. Moreover, as we
shall see, the Imry-Ma state necessarily contains singularities that make
its energy higher than that of the ordered state.

In the opposite case of $n-1>d$ the mapping of the ${\bf S}$-space
onto the ${\bf r}$-space that generates topological objects is impossible.
They are absent together with the energy barriers and pinning.
The stable state of the system is unique and independent of the initial
condition. In this case the long-range order is destroyed in a manner
that agrees quantitatively with the Imry-Ma picture. This applies
to the Heisenberg model with $n=3$ (and greater) in one dimension,
$n=4$ (and greater) in two dimensions, and $n=5$ (and greater) in
3$d$. 

The case of $n-1=d$ is the borderline between the above two cases.
It corresponds to non-singular topological objects: Kinks in the $xy$
model in 1$d$, skyrmions in the Heisenberg model with $n=3$ in 2$d$,
and similar non-singular solutions for $n=4$ in 3$d$. They are characterized by
the topological charge. Its conservation is important as it is only weakly violated by the discreteness of 
the lattice and weak random field. Possession of a pinned topological charge by 
the Imry-Ma state prevents the system from  relaxing to this state from any initial 
state that has a different topological charge. 

\begin{figure}
\begin{centering}
\includegraphics[width=8cm]{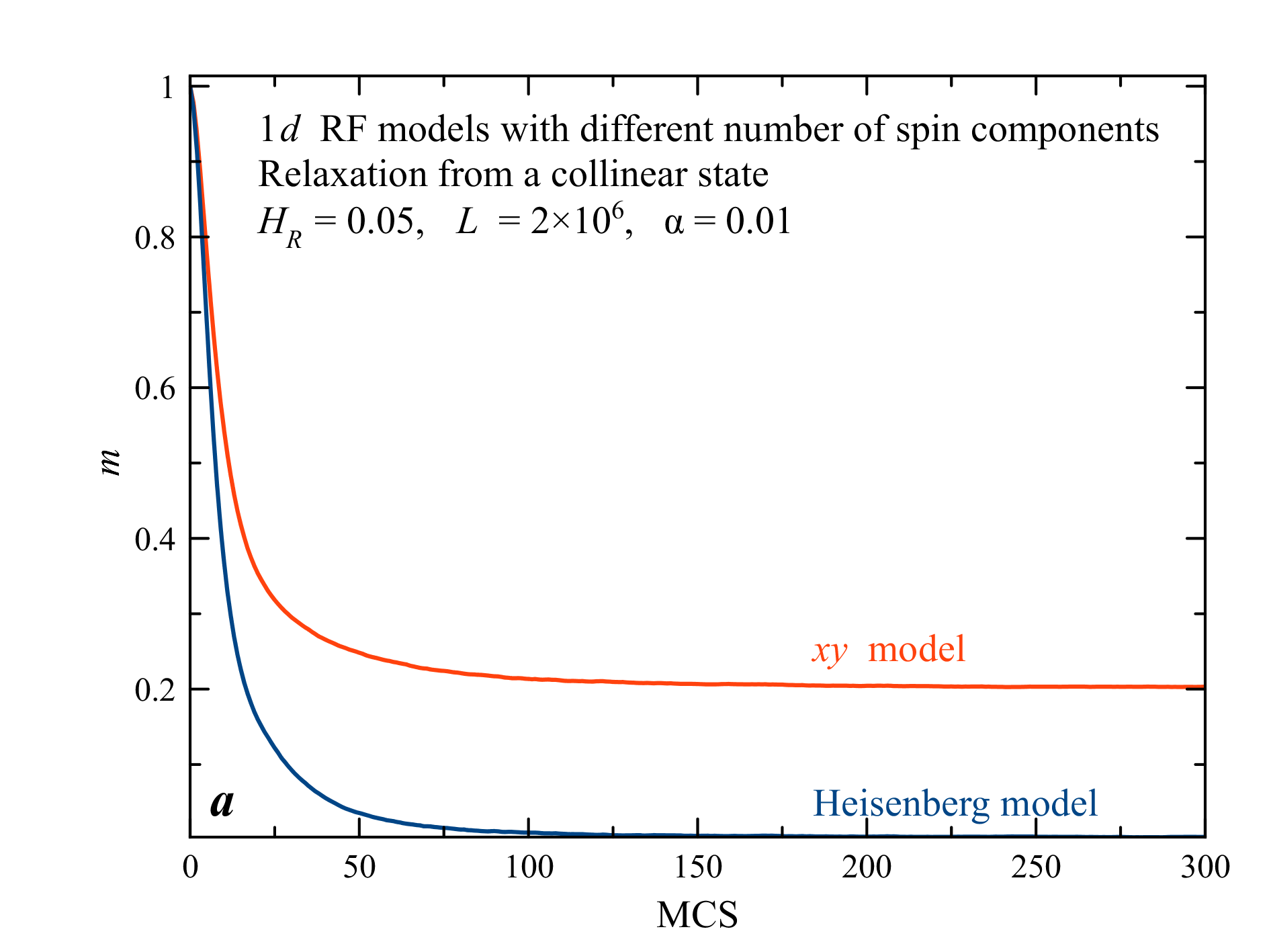}
\includegraphics[width=8cm]{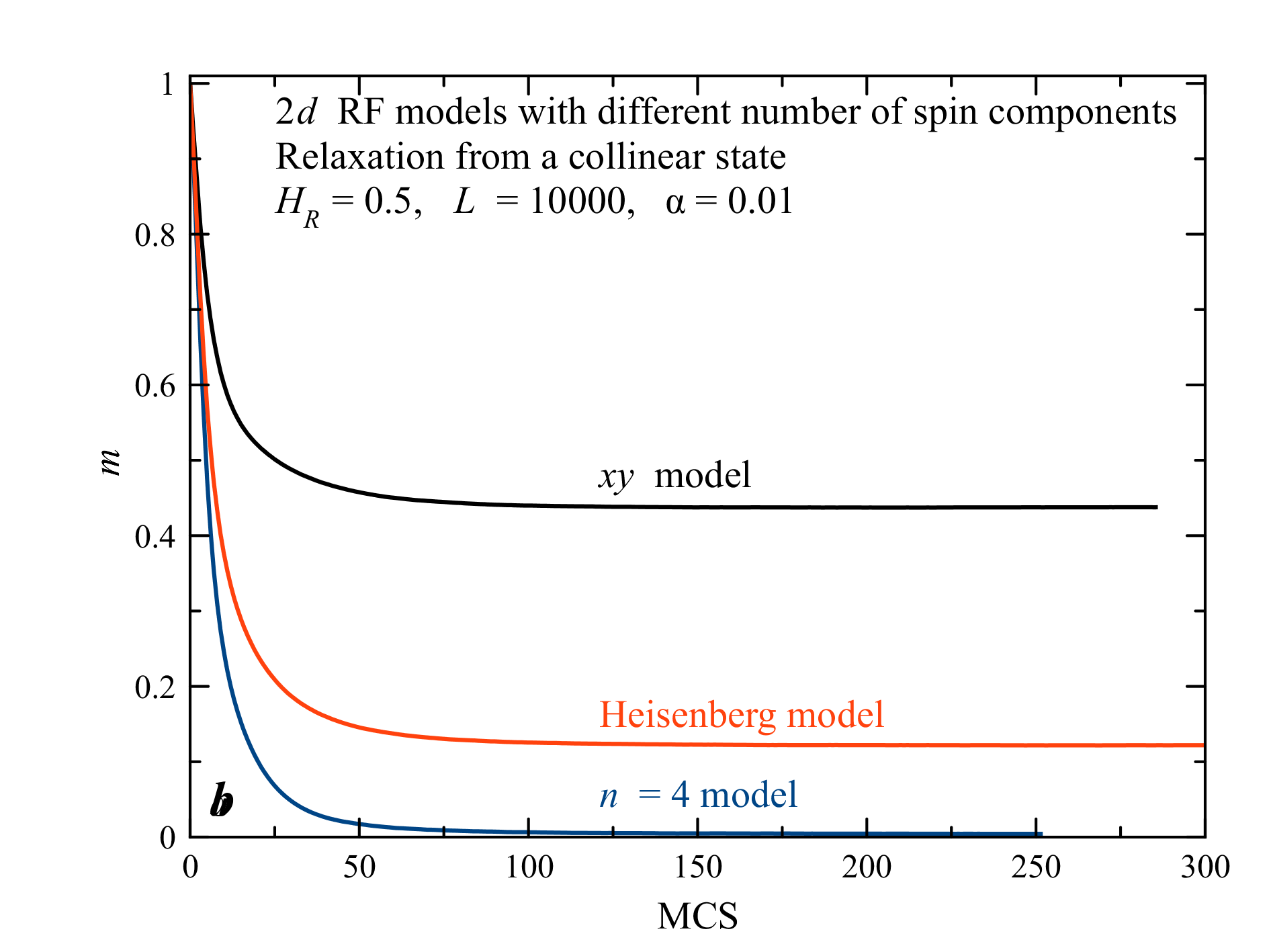}
\includegraphics[width=8cm]{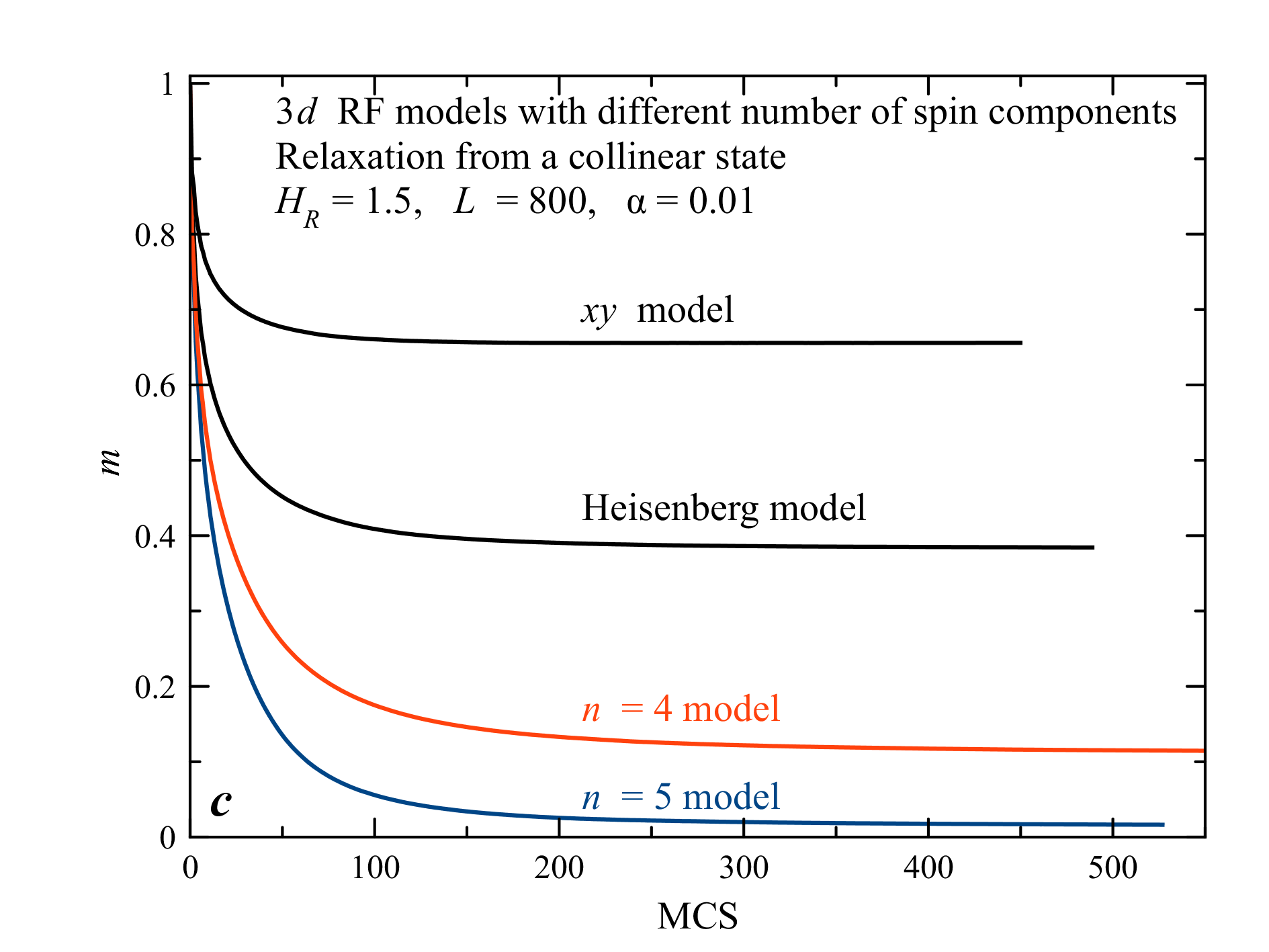} 
\par\end{centering}

\caption{Relaxation of the magnetization of the random-field spin system from
fully ordered initial state for different $d$ and $n$: (a) $d=1$,
$n=2,3$; (b) $d=2,$ $n=2,3,4$; (c) $d=3$, $n=2,3,4,5$. MCS means
a full spin update, as in Monte Carlo simulations.}

\label{Fig_m_vs_MSC} 
\end{figure}

To illustrate the validity of the above arguments, we have numerically studied
the discrete counterpart of the Hamiltonian (\ref{Ham-continuous})
with the nearest-neighbor exchange interaction in the presence of
the external field ${\bf H}$, 
\begin{equation}
{\cal H}=-\frac{1}{2}\sum_{ij}J_{ij}{\bf s}_{i}\cdot{\bf s}_{j}-\sum_{i}{\bf h}_{i}\cdot{\bf s}_{i}-{\bf H}\cdot\sum_{i}{\bf s}_{i},\label{Ham-descrete}
\end{equation}
on lattices containing hundreds of millions spins ${\bf s}_{i}$ of length $s$. The relation between parameters of the continuous and discrete models is $\alpha = Ja^{d+2}$, $S_0 = s/a^d$, where $a$ is the lattice parameter. We consider hypercubic lattices with periodic boundary conditions containing $L^d$ spins; $L$ being the linear size of the system. In computations we use $J=s=a=1$ and $h=H_R$.  Our numerical method combines sequential rotations of spins towards
the direction of the local effective field, ${\bf H}_{i,{\rm eff}}=\sum_{i}J_{ij}{\bf s}_{j}+{\bf h}_{j}+{\bf H}$,
with energy-conserving spin flips: ${\bf s}_{i}\rightarrow2({\bf s}_{i}\cdot{\bf H}_{i,{\rm eff}}){\bf H}_{i,{\rm eff}}/H_{i,{\rm eff}}^{2}-{\bf s}_{i}$
(so-called overrelaxation), applied with probabilities $\alpha$ and
$1-\alpha$ respectively; $\alpha$ playing the role of the relaxation
constant. Refs. \onlinecite{GCP-EPL13,GCP-PRB13} show high efficiency
of this method for glassy systems under the condition $\alpha\ll1$ which is physically equivalent to slow cooling. Relaxation of the per-site magnetization, $m=\sqrt{\mathbf{{m}\cdot{m}}}$,
where ${\mathbf{m}}=(sN)^{-1}\sum_{i}\mathbf{S}_{i}$, out of
a collinear state is shown for different $d$ and $n$ in Fig. \ref{Fig_m_vs_MSC}.

In one dimension, $n=2$ is the marginal case. Numerical analysis of
different spin configurations shows that the Imry-Ma-like state with
$m=0$ has the lowest energy. This state, however, cannot be achieved
through relaxation from the initially ordered state without forming
non-singular kinks or antikinks associated with the full clockwise
or counterclockwise rotations of the spin as one moves along the spin chain. 
They become pinned by the random field. The difference in the number of kinks and antikinks
is a conserved topological charge. While the system tends to disorder, it cannot do
so completely because it requires changing the topological charge. However, for three-component
spins in one dimension, topologically stable objects are absent and
the system disorders completely as is illustrated by Fig. \ref{Fig_m_vs_MSC}a.

Two-component spins in two dimensions form well-known topological
singularities -- vortices in the $xy$ model \cite{CT-book,Dieny-PRB1990,GCP-PRB13}.
Here again the system wants to relax to the Imry-Ma-like state with
$m=0$ but cannot do it without forming vortices that cost energy,
which explains the curve in Fig. \ref{Fig_m_vs_MSC}b for the $xy$
model in 2$d$.

In the marginal case of $d=2$, $n=3$ the model possesses non-singular
topological objects -- skyrmions \cite{CT-book}. In the absence of
the random field the difference in the number of skyrmions and antiskyrmions
is a conserved topological charge. Skyrmions on the lattice tend to
collapse \cite{skyrmion}. However, pinning by the random field stabilizes
them. We have checked numerically that for $d=2$, $n=3$ the Imry-Ma
state with $m=0$ has the lowest energy. However, conservation of
the topological charge prevents the system from relaxing to this
state from almost any initial condition. This effect is responsible
for a small but finite magnetization obtained by the relaxation from the initially
ordered state, see Fig. \ref{Fig_m_vs_MSC}b. However, for
a four-component spin in two dimensions, topological objects are absent and 
the system relaxes to the state with a zero magnetic moment, see Fig.
\ref{Fig_m_vs_MSC}b.

\begin{figure}
\begin{centering}
\includegraphics[width=7cm]{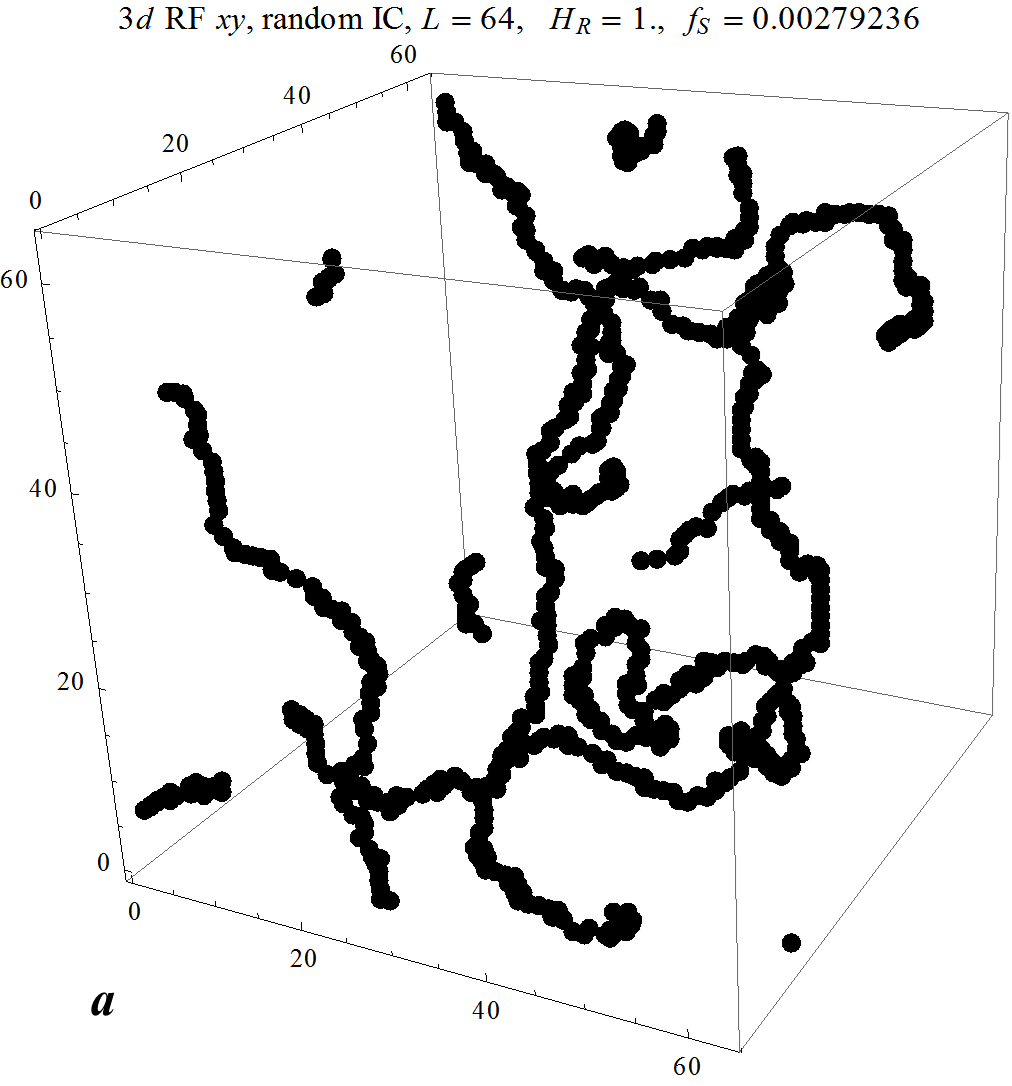}
\includegraphics[width=7cm]{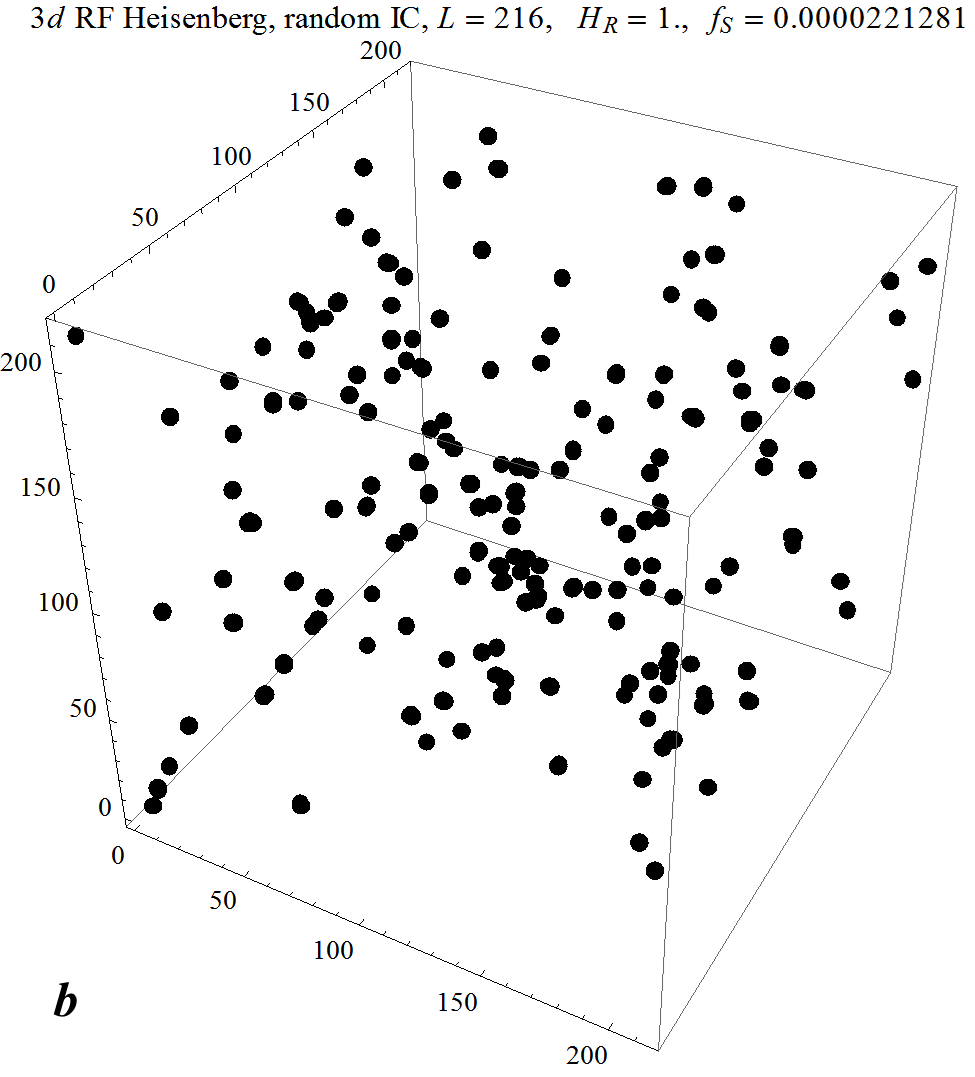} 
\par\end{centering}
\caption{Topological simgularities in the random-field spin model in three
dimensions obtained by relaxation from random initial orientation
of spins: (a) Pinned vortex loops of the $xy$ $(n=2)$ model; (b)
Pinned hedgehogs of the Heisenberg $(n=3)$ model. $f_{S}$ is fraction
of the lattice interstitial (body centered) sites occupied by singularities.}
\label{loops} 
\end{figure}

Relaxation in a three-dimensional case is illustrated by Fig. \ref{Fig_m_vs_MSC}c.
For $n=2$ the system possesses vortex lines or loops that in the
lattice model are singular pancake vortices in $2d$ planes stuck
together, see Fig. \ref{loops}a.
Similarly, the model with three-component spins in 3$d$ has singular
hedgehogs, see Fig. \ref{loops}b. The energy cost of vortex loops and hedgehogs
prevents the 3$d$ system of spins from relaxing to the $m=0$ state,
as is shown in Fig. \ref{Fig_m_vs_MSC}c. Starting from random orientation of spins one obtains states of vortex or hedgehog glasses with $m=0$ and energies higher than those of the ordered states. 

In the marginal case of $n=4$ the $3d$ random-field model has non-singular
topological structures pinned by the random field which are similar
to skyrmions in $2d$. In this case the final magnetic moment is still
non-zero but small, see Fig. \ref{Fig_m_vs_MSC}c. We again find that
the energy of the Imry-Ma-like state $m=0$ state for $d=3$, $n=4$
is lower than that of the $m\neq0$ state. However, the difference
in the topological charge prevents the system from relaxing to the
Imry-Ma state from almost any initial state.

The model with five-component spins in 3$d$ does not possess any
topologically stable structures. The relaxation of the system from
the ordered initial state is unobstructed by any topological arguments
and the system ends up in the state with $m=0$, Fig. \ref{Fig_m_vs_MSC}c.

\begin{figure}
\begin{centering}
\includegraphics[width=8cm]{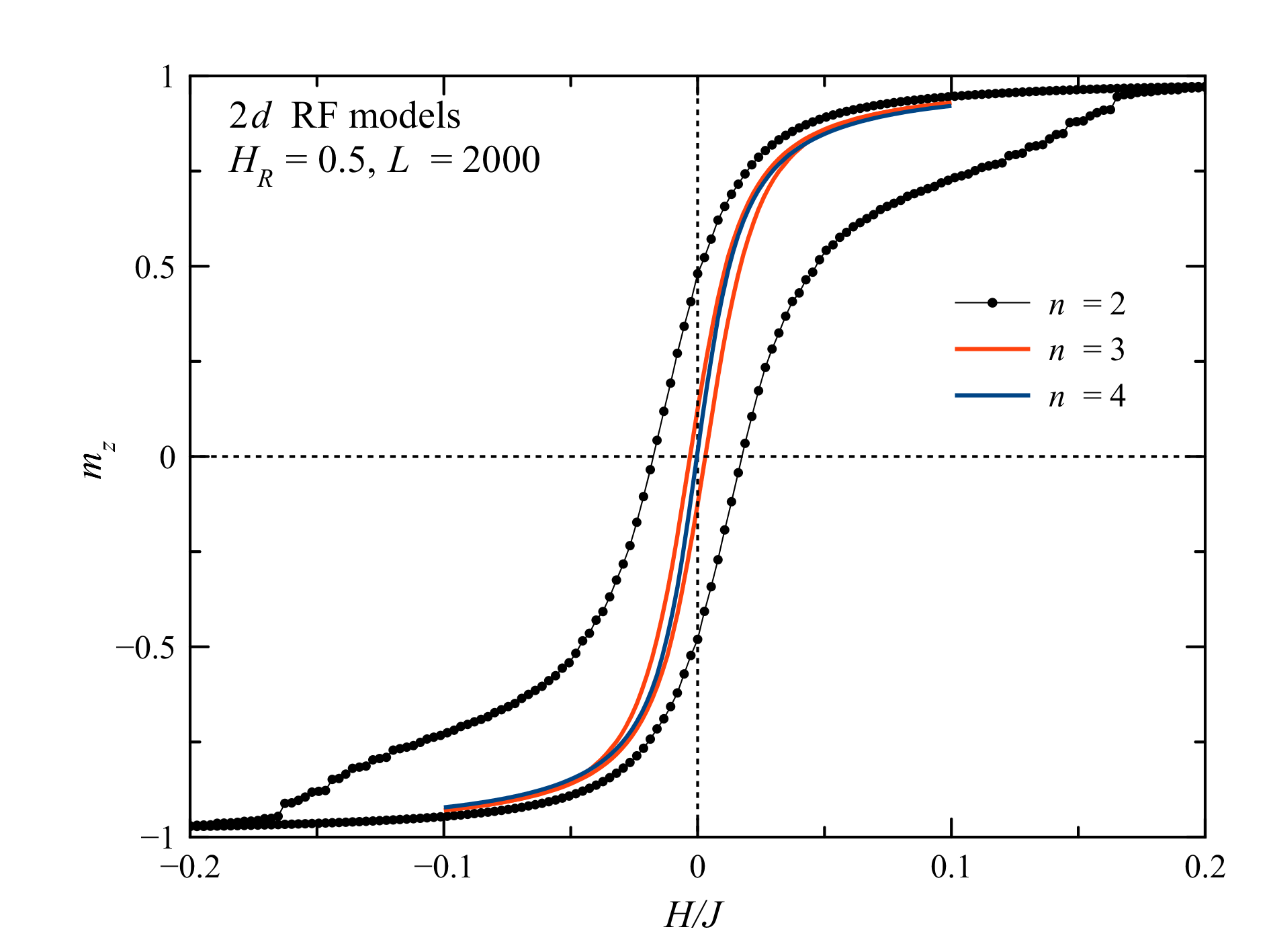} 
\par\end{centering}
\caption{Hysteresis curves of the random field spin model in two dimensions
for $n=2,3,4$.}
\label{hysteresis} 
\end{figure}

The relation between topology and metastability in, e.g., two spatial
dimensions is further illustrated by the hysteresis curves in Fig.
\ref{hysteresis}. The model with $n=2$, that possesses $xy$ vortices
with singularities, is characterized by sizable hysteresis loop which
is indicative of strong metastability. The loop becomes thin for $n=3$
when non-singular skyrmions are present. It disappears completely,
resulting in a reversible magnetic behavior, at $n=4$ when topological
objects are absent. Similar behavior for different $n$ has been observed
in 3$d$.

To get a better understanding of how topology modifies the Imry-Ma argument and leads to metastability let us recall that in that argument the order parameter ${\bf S}$ follows the direction of the average random field, $\bar{\bf h}$, on a scale $R_{f}$. The energy of the random field in Eq. (\ref{Ham-continuous}) is proportional to $-\bar{h}S_0\sim-hS_0/R_{f}^{d/2}$,
while the non-uniformity energy is proportional to $\alpha S_0^2/R_{f}^{2}$.
Minimization of the total energy with respect to $R_{f}$ then leads to the rotation of ${\bf S}$ by a significant angle on a scale $R_{f}\propto(Js/h)^{2/(4-d)}$. For $R\gtrsim R_{f}$ correlations should be completely destroyed,
thus the state of the system should be disordered. This famous argument, however, does not account for the energy associated with unavoidable singularities at $n - 1 < d$. 

To show this, consider
components of the averaged random field $\bar{h}_{\alpha}$, $\alpha=1,\ldots,n$.
Since $\bar{h}_{\alpha}$ are sums of many random numbers, they are
statistically independent and have Gaussian distribution. In about
a half of the space $\bar{h}_{\alpha}>0$, in the other half $\bar{h}_{\alpha}<0$.
Boundaries between these regions are subspaces of dimension $d-1$,
where $\bar{h}_{\alpha}=0$. Intersection of all these subspaces,
that is, $\mathbf{\bar{h}}=0,$ is unavoidable and forms a subspace
of dimension $d-n$, if $n\leq d$. It is easy to see that subspaces with $\mathbf{\bar{h}}=0$
are singularities in the spin field $\mathbf{S}$. Since $\mathbf{S}^{2}=\mathrm{const}$,
crossing subspaces $\mathbf{\bar{h}}=0$ makes all components of $\mathbf{S}$
change the direction. 

\begin{figure}
\begin{centering}
\includegraphics[width=6.5cm]{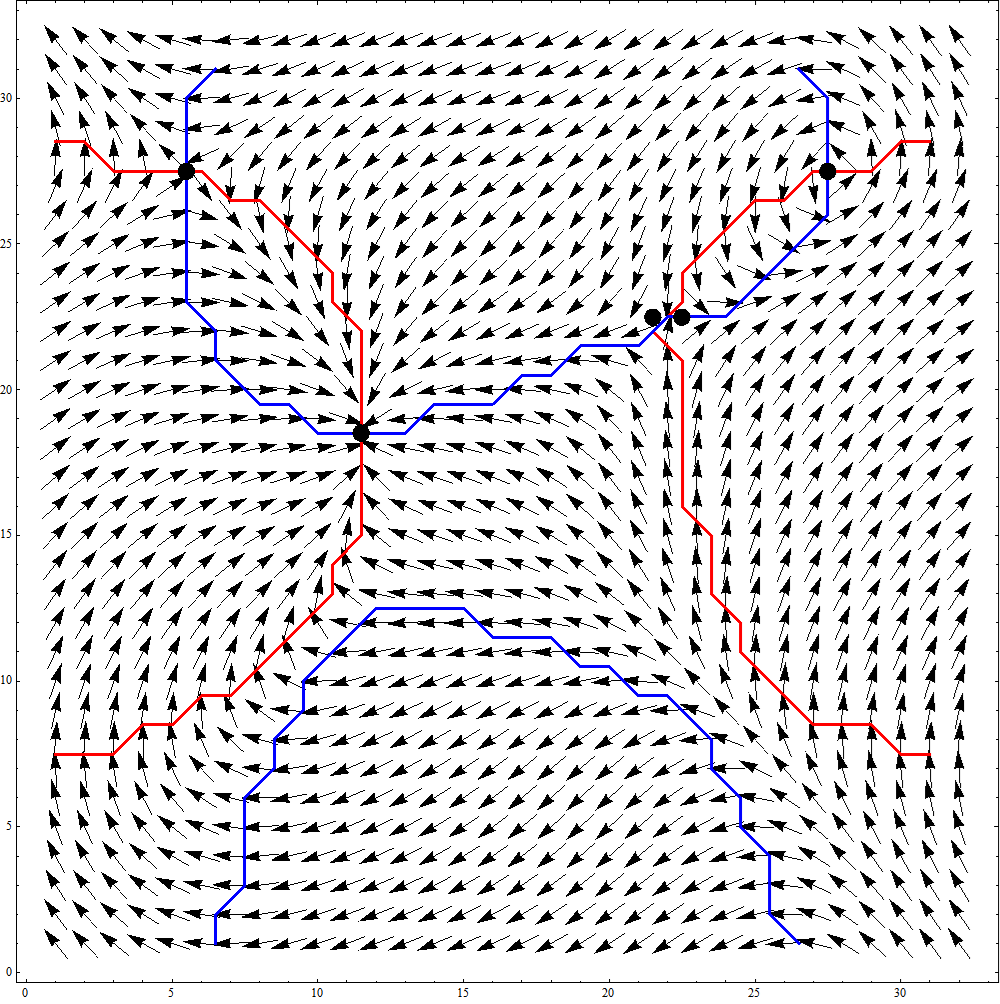} 
\par\end{centering}
\caption{Emergence of vortices and antivortices at the intersections of lines corresponding to $\bar h_x = 0$ (red) and $\bar h_y = 0$ (blue) in the random-field $2d$ $xy$ model. Picture reflects numerical averaging of ${\bf h}$ within finite range according to the prescription of the Imry-Ma model for a particular realization of the random field. Arrows show spins on lattice sites. Similar structures emerge after relaxation from random orientation of spins, with the positions of vortices depending on the initial state.}
\label{Fig-IM-spins-2d} 
\end{figure}

\begin{figure}
\begin{centering}
\includegraphics[width=7cm]{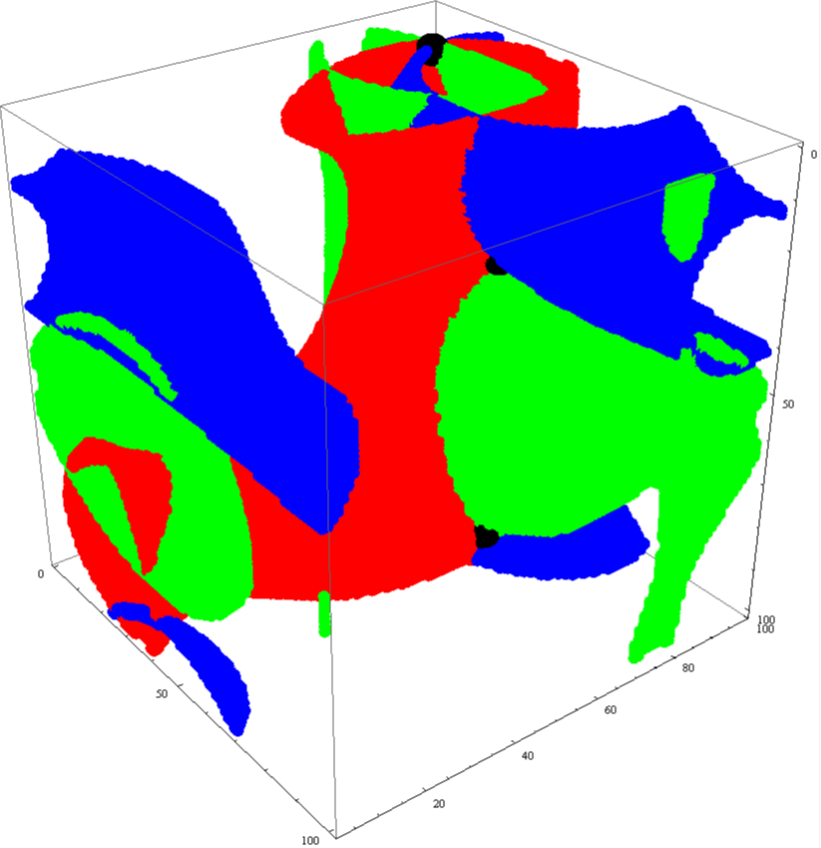} 
\par\end{centering}
\caption{Emergence of hedgehogs (black) at the intersection of three surfaces (shown in different color) corresponding to $\bar h_x = 0$, $\bar h_y = 0$, and $\bar h_z = 0$ respectively in the $n = 3$ random-field model in three dimensions.}
\label{spheres} 
\end{figure}

For $n=2$ in $2d$ subspaces $\mathbf{\bar{h}}=0$ are points and the corresponding singularities are vortices or antivortices. A spin field in the $2d$ $xy$ model generated in accordance with the Imry-Ma prescription is shown in Fig. \ref{Fig-IM-spins-2d}. The red line corresponds to $\bar h_x = 0$ and thus spins directed along the $y$-axis. The blue line corresponds to $\bar h_y = 0$ and thus spins directed along the $x$-axis. At the intersections of red and blue lines the spins can look neither in the $x$ nor in the $y$-direction. This generates topological defects -- vortices or antivortices.
For $n=2$ in 3$d$ subspaces $\mathbf{\bar{h}}=0$ are lines and
the singularities are vortex lines or loops. For $n=3$ in 3$d$ subspaces
$\mathbf{\bar{h}}=0$ are points and the singularities are hedgehogs. They emerge at  the intersection of surfaces corresponding to $\bar h_x = 0$, $\bar h_x = 0$, and $\bar h_x = 0$, see Fig. \ref{spheres}.
Singularities push the energy of the Imry-Ma state up and (according
to our numerical results) make it higher than the energy of the ordered
state. However, for $n>d$, the averaged random field is
non-zero everywhere and the spin field $\mathbf{S}$ is non-singular. Consequently,
at $n>d$ the $m = 0$ state has the lowest energy in accordance with the Imry-Ma argument and the Aizenman-Wehr theorem that assume continuity of the spin field. Still at $n = d+1$ the presence of non-singular topological objects and conservation of topological charge prevents the ordered state from relaxing to the $m =0$ Imry-Ma state. Only at 
$n > d+1$, when the spin-field is continuous and topological objects are absent the system relaxes to the Imry-Ma state. 

\begin{figure}
\begin{centering}
\includegraphics[width=8cm]{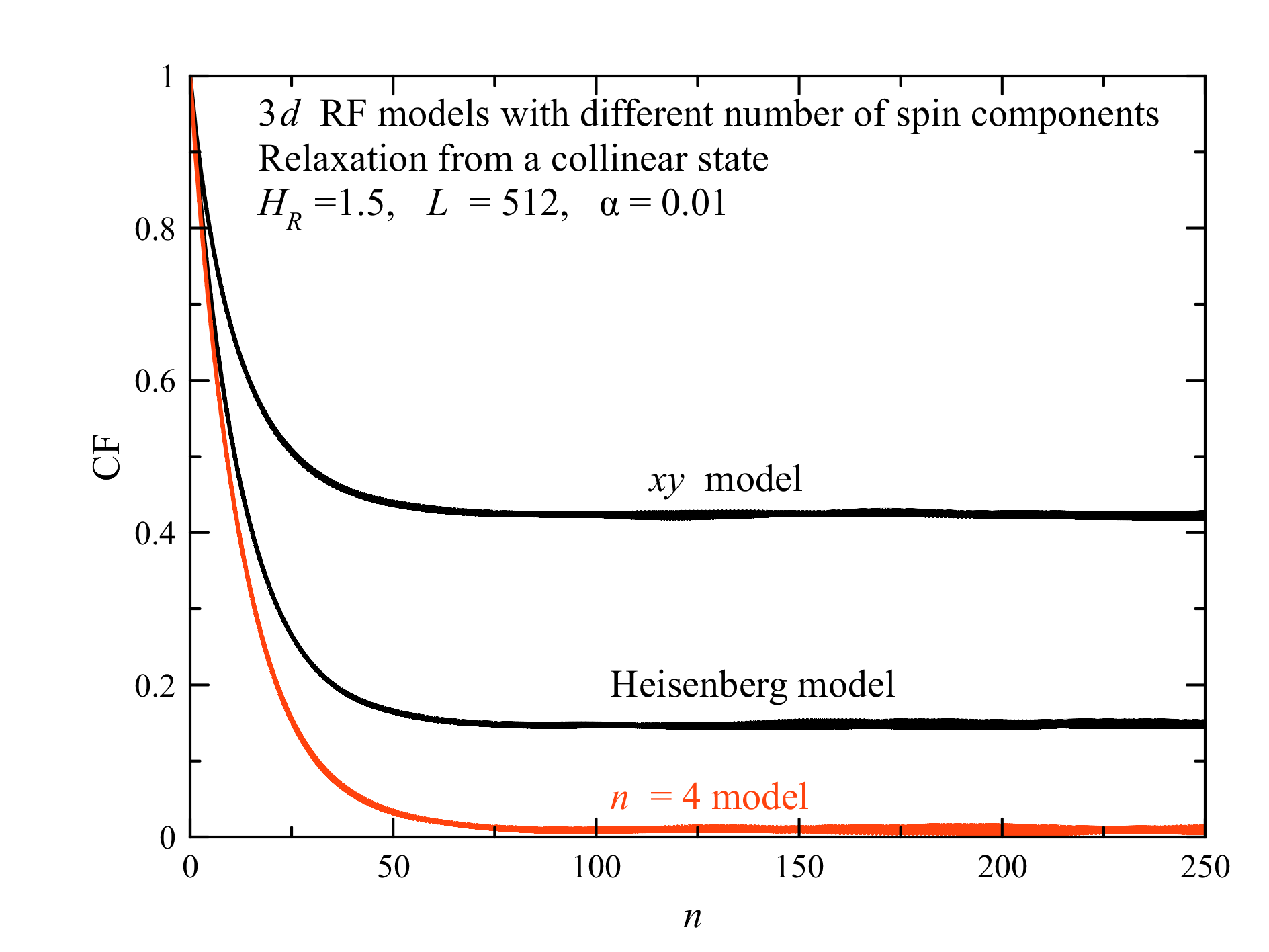} 
\par\end{centering}
\caption{Spin-spin correlation function of the random-field model for $n=2,3,4$
in three dimensions.}
\label{CF-3d} 
\end{figure}
\begin{figure}
\begin{centering}
\includegraphics[width=7.4cm]{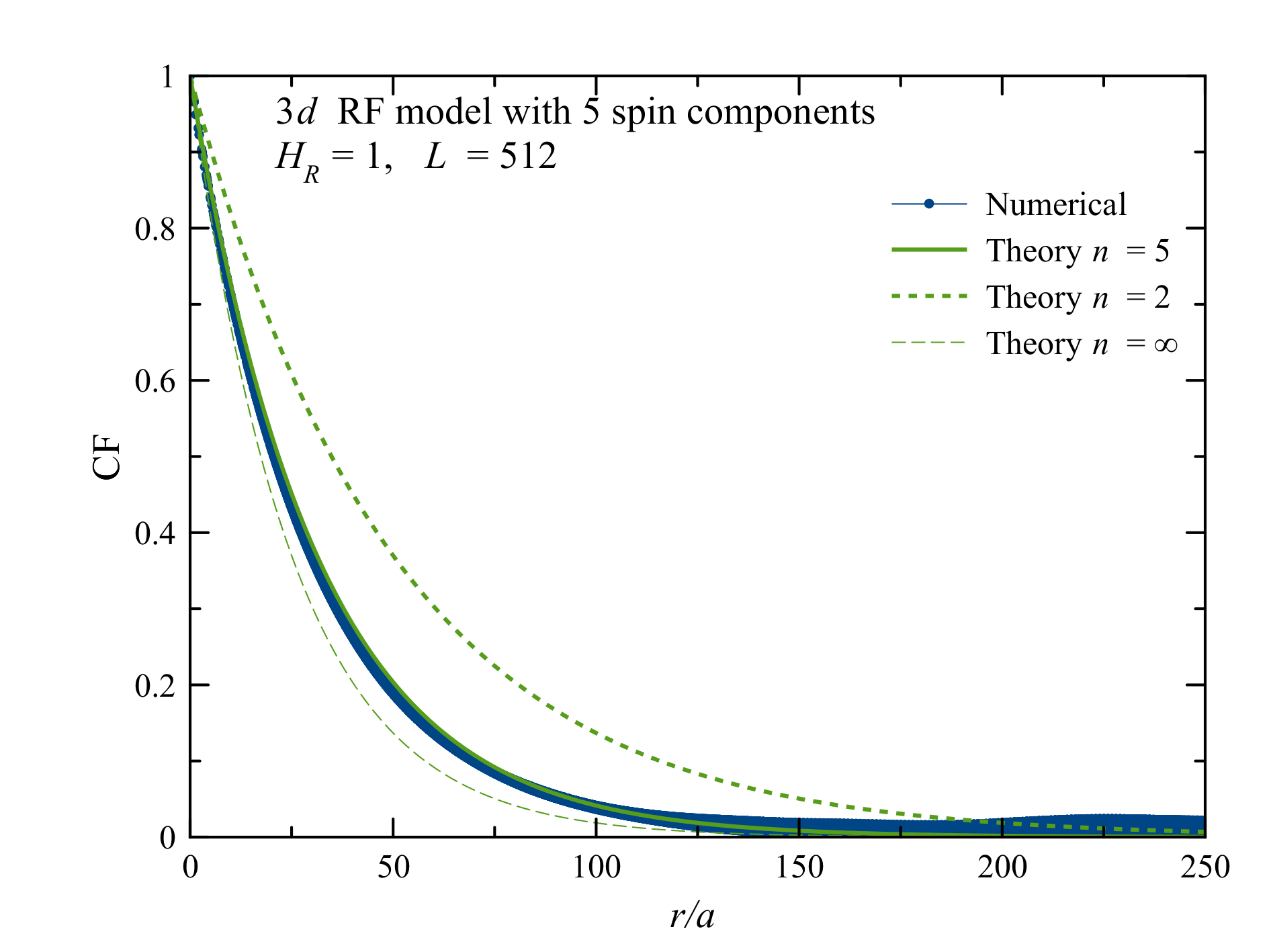} 
\par\end{centering}
\caption{Theoretical and numerical spin-spin correlation functions of the $n=5$
random-field model in three dimensions.}
\label{CF-5} 
\end{figure}

To get a better idea of the long-range correlations in the random-field model we have computed the spin-spin correlation function in the final state obtained through relaxation from the initially ordered state. Fig. \ref{CF-3d} shows the $3d$ spin-spin correlation function for
$n=2,3,4$. Note that the height of the plateau at large distances equals $m^2$ for $R_f \ll L$. For $n=2,3$ the ferromagnetic order persists at all
distances. For $n=4$ the long-range correlation plateau
becomes very low in accordance with Fig. \ref{Fig_m_vs_MSC}c. 
The $3d$ correlation function for $n=5$ is shown in Fig. \ref{CF-5}. Here, in accordance with the Imry-Ma picture, correlations are fully destroyed at large distances.

\begin{figure}
\begin{centering}
\includegraphics[width=8cm]{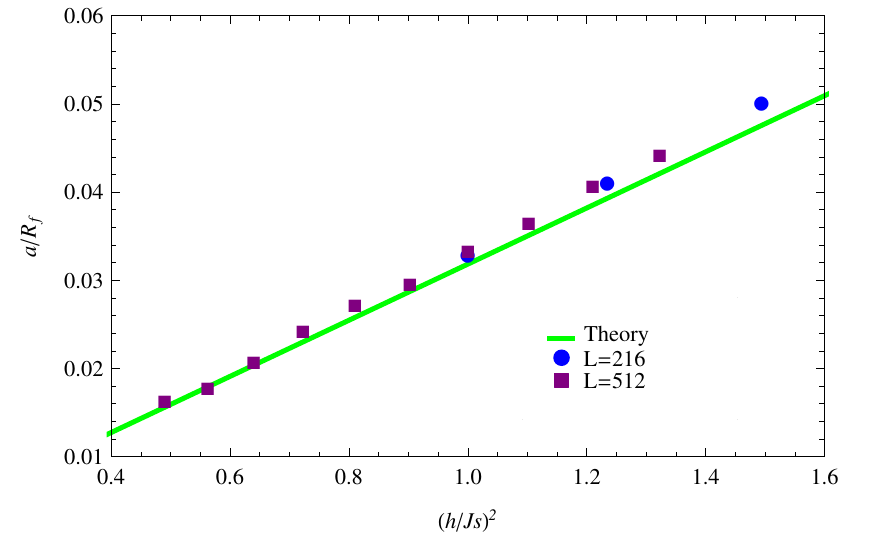} 
\par\end{centering}
\caption{Dependence of $R_f$ on $h$ computed numerically (points) for the $n = 5$ random-field model in three dimensions and given by Eq.\ (\ref{Rf}) (solid line) at $n=5$.}
\label{Rf-h} 
\end{figure}
We have analytically derived from Eq. (\ref{Ham-continuous}) the
short-distance form of the spin-spin correlation function in 3$d$, $\langle{\bf s}({\bf r}_{1})\cdot{\bf s}({\bf r}_{2})\rangle\cong1-|{\bf r}_{1}-{\bf r}_{2}|/R_{f}$,
where 
\begin{equation}\label{Rf}
\frac{R_{f}}{a}=\frac{8\pi\alpha^{2}S_0^{2}}{h^{2}a^{4}(1-1/n)}=\frac{8\pi}{(1-1/n)}\left(\frac{Js}{h}\right)^{2}.
\end{equation}
In fact, this short-range form of the correlation function agrees with our numerical results for all $n$ in $3d$. For $n\geq5$ the spin-spin correlation function at all distances can be very well fitted by $\langle{\bf s}({\bf r}_{1})\cdot{\bf s}({\bf r}_{2})\rangle=\exp\left(-|{\bf r}_{1}-{\bf r}_{2}|/R_{f}\right)$
that we will loosely call ``theoretical''. A good agreement with
this formula is illustrated by Fig. \ref{CF-5}. So far we have been
able to prove analytically the numerically confirmed exponential decay
of the correlation function in $3d$ only for the mean-spherical model
which corresponds to $n=\infty$ \cite{Stanley}. However, the observed exponential behavior of  $\langle{\bf s}({\bf r}_{1})\cdot{\bf s}({\bf r}_{2})\rangle$ and the observed $1/h^2$ dependence of $R_f$ on the strength of the random field for small $h$ at $n = 5, d = 3$ (see Fig. \ref{Rf-h}) present clear evidence of the onset of the Imry-Ma state in the absence of topological objects.

In conclusion, we have demonstrated that the long-range order in random-field
systems is controlled by topology. For the $n$-component spin in
$d$ dimensions the presence of topological structures at $n-1\leq d$
invalidates the Imry-Ma argument which suggests destruction of the
long-range order by the arbitrary weak random field. For $n-1>d$, when
topological structures are absent, the long-range behavior of the
system agrees quantitatively with the Imry-Ma picture. This finding
solves the controversy that existed in the field for over four decades.
It provides the guiding principle for assessing the long-range behavior
of various systems with quenched randomness and continuous-symmetry
order parameter. 

This work has been supported by the Department of Energy through grant
No. DE-FG02-93ER45487.

\end{document}